\documentclass[prd,preprint,eqsecnum,nofootinbib,amsmath,amssymb,preprintnumbers,tightenlines,longbibliography,hyperlink]{revtex4-1}

\usepackage{graphicx}% Include figure files
\usepackage{dcolumn}% Align table columns on decimal point
\usepackage{bm}% bold math
\usepackage[colorlinks=true,linktocpage=true,linkcolor=blue,citecolor=blue]{hyperref}
\def\e{{\varepsilon} }

\def\eq{{\rm eq}}
\def\p{{\boldsymbol p}}
\def\q{{\boldsymbol q}}

\def\Eq#1{Eq.~(\ref{#1})}
\def\Eqs#1{Eqs.~(\ref{#1})}

\def\Fig#1{Fig.~\ref{#1}}
\def\Sect#1{Section~\ref{#1}}

\def\tr{{\rm tr}}

\newcommand \beq{\begin{eqnarray}}
\newcommand \eeq{\end{eqnarray}}
\newcommand \be{\begin{equation}}
\newcommand \ee{\end{equation}}

\begin{document}

%\preprint{APS/123-QED}

\title{The weak magnetic field effect on dilepton polarization in heavy-ion collisions} % and dilepton anisotropy}% Force line breaks with \\

\author{Minghua Wei}
\author{Li Yan}
\affiliation{Institute of Modern Physics, Fudan
University, 220 Handan Road, Shanghai 200433, China}
\affiliation{Key Laboratory of Nuclear Physics and Ion-beam Application (MOE), Fudan University, Shanghai 200433, China}

\date{\today}% It is always \today, today,

\begin{abstract}

The measurement of the magnetic field created in high-energy heavy-ion collisions is challenging, due the the fact that the magnetic field decays so drastically that in a thermalized quark-gluon plasma the field strength becomes rather weak. By incorporating the weak magnetic effect into the medium, and especially into the production formalism of dileptons from the quark-gluon plasma, the effect of dilepton polarization is studied through the dilepton angular distribution. We find that the anisotropic coefficients in the dilepton spectrum are quite sensitive to the orientation and strength of the weak field. Accordingly, these coefficients provide ideal probes for the magnetic field in realistic experiments.

\end{abstract}

\maketitle

\section{Introduction}

A super strong electromagnetic field is created in the high-energy heavy-ion collisions. Through simple arguments from classical electrodynamics, the strength of the magnetic field is expected up to $10^{19}$ Gauss at the top energies at the Relativistic Heavy-Ion Collider at the Brookhaven National Laboratory~\cite{Kharzeev:2007jp,Skokov:2009qp,Deng:2012pc}. However, experimental signatures associated with the electromagnetic field are by far rare~\cite{Muller:2018ibh, STAR:2023nvo, STAR:2023jdd}, due to, to a large extent, the fact that the field decays so drastically that it becomes rather weak in thermalized quark-gluon plasma (QGP)~\cite{Kharzeev:2007jp,Skokov:2009qp, Deng:2012pc, Huang:2022qdn}.

To incorporate the weak magnetic effect in a theoretical description of the QGP evolution, especially in hydrodynamical modeling which has been found extremely successful~\cite{Shen:2020mgh}, % with respect to many observables,  
a natural way is to generalize the dissipative correction by including a finite electrical current (cf. \cite{DeGroot:1980dk}). Such a current leads to effectively a weak coupling of the quarks in the QGP flow to the external electromagnetic field. As a consequence, the motion of quarks in a fluid cell is corrected according to the orientation and strength of the field, resulting in observables. In previous studies, the weak magnetic effect has been found responsible for the anisotropic emission of direct photons~\cite{Sun:2023pil, Sun:2023rhh}, and the sign change of the local lambda hydron polarization~\cite{Sun:2024isb}. 

In this study, we would like to investigate in a similar way, the weak magnetic effect on the dilepton polarization. Unlike the strong magnetic field assumption~\cite{Tuchin:2013bda, Sadooghi:2016jyf, Das:2019nzv, Wang:2020dsr,Chaudhuri:2021skc, Wang:2022jxx}, in our analysis, the strength of the magnetic field inside QGP is treated weak, subject to the condition that $|eB|\ll m_\pi^2$, with $m_\pi$ the pion mass. In high-energy heavy-ion experiments, dileptons provide a perfect electromagnetic probe for QGP. In addition to the cocktail contributions from hadron decays, heavy quarks, initial hard processes, etc., the fraction of the dilepton production from QGP thermal radiation is strongly correlated to the early-stage properties of the medium~\cite{Serreau:2003wr, Ding:2010ga}. The polarization of dileptons from QGP, or more precisely, the polarization of virtual photons, can be measured via the angular distribution of the dilepton spectrum~\cite{NA60:2008iqj, HADES:2011nqx, Speranza:2016tcg, Speranza:2018osi, Seck:2023oyt, Coquet:2023wjk}. The detailed framework of calculations will be given in \Sect{sec:formalism}. With respect to the Bjorken flow solution to the QGP expansion, with a weak magnetic field whose decay is controlled by a short lifetime, we calculate the anisotropic coefficients of the dilepton angular distribution in \Sect{sec:numericalresults}.  Interestingly, the existence of a weak magnetic field leads to unambiguous signatures of the polarization properties of the dileptons, regarding their emission azimuthal angle.

%\begin{equation}
%\begin{aligned}
%\frac{d R}{d^4 q d \Omega_{\ell}}= & \mathcal{N}\left(1+\lambda_\theta \cos ^2 \theta_{\ell}\right. \\
%& +\lambda_\phi \sin ^2 \theta_{\ell} \cos 2 \phi_{\ell}+\lambda_{\theta \phi} \sin 2 \theta_{\ell} \cos \phi_{\ell} \\
%& \left.+\lambda_\phi^{\perp} \sin ^2 \theta_{\ell} \sin 2 \phi_{\ell}+\lambda_{\theta \phi}^{\perp} \sin 2 \theta_{\ell} \sin \phi_{\ell}\right),
%\end{aligned}
%\end{equation}
%
%
%
%This manuscript is organized as follows: In Section \ref{sec:formalism}, we will introduce the electromagnetic effects on the quark distribution functions. Consequently, for $q\bar{q}\rightarrow l\bar{l}$ process, photon tensor $W^{\mu \nu}$ and dilepton production rate are modified by a magnetic field. In the helicity frame, anisotropic coefficients $\lambda_{\phi}$, $\lambda_{\theta}$ and $\lambda_{\theta\phi}$ is determined by photon polarization. In Section \ref{sec:numericalresults}, we will demonstrate the numerical results of $\lambda_{\phi}$ and $\lambda_{\theta}$ in a static fluid and Bjorken flow. Finally, Section.\ref{sec:conclusion} will provide the conclusion and outlook.
%
%Covention. We shall use captital letters for four-vectors, while three-vector will be denoted by lower case letters.

\section{Theoretical formulation}
\label{sec:formalism}

\subsection{Dilepton production rate in QGP}

In a QGP medium, thermal dileptons are those produced by a virtual photon from the annihilation of a quark and an antiquark: $q\bar q\to \gamma^*\to l\bar l$. The differential production rate per volume is well known from perturbative calculations~\cite{McLerran:1984ay,Weldon:1990iw}:
%\mh{Although Ref~\cite{McLerran:1984ay,Weldon:1990iw} omitted $\int$ as well, it's better to write down $\int$ like Ref~\cite{Speranza:2018osi}.  }
\begin{equation}
\label{eq:rate}
    \frac{d R}{d^4 Q}=\frac{e^4}{Q^4} \int W^{\mu \nu}(Q) L_{\mu \nu}(Q_1,Q_2) 
\delta^{(4)}\left(Q-Q_1-Q_2\right) \frac{d^3 \mathbf{q_1}}{(2 \pi)^3 2 E_{\mathbf{1}}} \frac{d^3 \mathbf{q_2}}{(2 \pi)^3 2 E_{2}}\,,
\end{equation}
where $Q$ is the four momentum of the virtual photon and it is normalized as the invariant mass, $Q^2=M^2>0$. In \Eq{eq:rate}, $L^{\mu\nu}$ is the lepton tensor.
Given the momentum of the lepton pair, $q_1$ and $q_2$, 
%are the momentum of lepton and anti-lepton. In principle, both $W^{\mu\nu}$ and $L_{\mu\nu}$ are affected by magnetized medium. Lepton pairs penetrate the QGP, evolve in the vacuum, and are detected as plane waves. For plane waves, 
after a summation over spins
the lepton tensor can be found as\footnote{
Note, however, that 
\Eq{eq:lmunu} corresponds to plane-wave solution which is obtained with respect to free lepton pairs. In principle, if a strong magnetic field is applied to the system, the formulation of the lepton tensor should be modified according to the correction to the motion of the lepton pairs inside the magnetic field, see Ref.~\cite{Sadooghi:2016jyf}.
}:
\begin{equation}
\label{eq:lmunu}
L^{\mu \nu}(Q_1,Q_2)=-4\left[
    (Q_1\cdot Q_2+m_{l}^{2})g^{\mu\nu}-Q_1^{\mu} Q_2^{\nu}-Q_1^{\nu}Q_2^{\mu}\right]\,,
\end{equation}
%where $q=l+\bar{l}$ and $k=l-\bar{l}$. We set $q^{2}=M^{2}$ and $M$ is the invariant mass.
where $m_{l}$ is the mass of the lepton. 
%We set $q^{2}=M^{2}$ and $M$ is the invariant mass. Then the integration of $\mathbf{l}$ and $\bar{\mathbf{l}}$ can be calculated as follows:
%\begin{equation}
%\int \frac{d^3 \mathbf{l}}{E_{\mathbf{l}}} \int \frac{d^3 \bar{\mathbf{l}}}{E_{\bar{\mathbf{l}}}} \delta^4\left(l+\bar{l}-q\right) L_{\mu \nu}\left(l,\bar{l}\right)=\frac{2 \pi}{3} \left(1+\frac{2 m_{l}^2}{M^2}\right)\left(1-\frac{4 m_{l}^2}{M^2}\right)^{1 / 2}\left(q_\mu q_\nu-M^2 g_{\mu \nu}\right)
%\end{equation}

The photon tensor, $W^{\mu\nu}$, originated from the in-medium correlation of electromagnetic current. To the lowest order in the electromagnetic coupling $e$, but to all orders in the strong coupling constant $g$, $W^{\mu\nu}$ is related to the one-particle irreducible photon self-energy, $\Pi^{\mu\nu}$. The photon tensor contains the information of %the virtial photon from %, and in particular, 
the QGP medium, where the dilepton is produced.  
%on the other hand, should be averaged with respect to the QGP medium. 
In a kinetic theory approach, regardless of whether the QGP is in local thermal equilibrium, one has %taking into account the effect of out-of-equilibrium corrections, one has
%In Ref.\cite{Speranza:2018osi}, electromagnetic current correlation function $W^{\mu \nu}$ in finite temperature is obtained by ensemble average:
\begin{equation}
W^{\mu \nu}=\left\langle w^{\mu \nu}\right\rangle\,,
\end{equation}
where the angular bracket $\left\langle ... \right\rangle$ denotes an average with respect to the quark phase-space distribution $f_q$ and $f_{\bar q}$,
\begin{equation}
\label{eq:emsemble}
\begin{aligned}
\langle A\rangle= & \int \frac{d^3 \p_1}{(2 \pi)^3 2 E_1} \frac{d^3 \p_2}{(2 \pi)^3 2 E_2}(2 \pi)^4 \delta^{(4)}\left(Q-P_1-P_2\right) f_{q} f_{\bar{q}} A .
\end{aligned}
\end{equation}
For a QGP in local equilibrium, the quark distribution function takes the form of the Fermi-Dirac distribution and depends explicitly on $P\cdot u/T$, with $u^\mu$ the fluid four velocity. For a QGP medium slightly out of local equilibrium due to dissipation, corrections to the distributions can be found correspondingly~\cite{Sun:2023rhh}. 
Especially, a weak electromagnetic field induces dissipative corrections which breaks azimuthal symmetry, irrespective of the medium expansion.  
%where $p_{1}=(E_{1},\mathbf{p}_1)$ and $p_{2}=(E_{2},\mathbf{p}_2)$ are the momenta of quark and anti-quark respectively. Here, we try to apply the distribution function $f_{q}\left(\mathbf{p}_1\right)$ in a magnetic field, rather than to obtain an electromagnetic current correlation function $W^{\mu \nu}(B)$ strictly. In Eq.\eqref{eq:emsemble}, we used $f_{q}\sim n_{\mathrm{eq}}+f_{\mathrm{EM}}$ where $n_{\mathrm{eq}}=1/(e^{p\cdot u}+1)$ and $f_{\mathrm{EM}}$ was given by Eq.\eqref{eq:fem}. This ansatz makes sure that $W^{\mu\nu}$ is a Lorentz-covariant function of $u^{\mu}$, $q^{\mu}$ and $F^{\mu\nu}$.
For the $q\bar{q}\rightarrow l\bar{l}$ process in QGP, one also has: 
%\ly{Need to be verified.}\textcolor{blue}{MH: the right-hand side is equal to $-4 C_q\left[-(P_{1}\cdot P_{2}+m_{q}^{2}) g^{\mu \nu}-P_{1}^{\mu} P_{2}^{\nu}-P_{2}^{\mu}P_{1}^{\nu}\right]$, which is related to $m_{q}$.}
\begin{equation}
w^{\mu \nu}=-4 C_q\left[-(P_{1}\cdot P_{2}+m_{q}^{2}) g^{\mu \nu}-P_{1}^{\mu} P_{2}^{\nu}-P_{2}^{\mu}P_{1}^{\nu}\right]
%2 C_q\left(-P^2 g^{\mu \nu}+P^\mu P^\nu-\Delta P^\mu \Delta P^\nu\right)
\end{equation}
where %$P=P_{1}+P_{2}$, $\Delta P=P_{1}-P_{2}$ and 
$C_q=5/3$ is a constant factor arising from the summation over quark flavors and colors.

\subsection{Dilepton production in terms of polarization states}

Similar to the decay of a vector meson, where the polarization states of a meson can be detected in terms of the daughter particles' angular distribution~\cite{Liang:2004xn, ALICE:2020iev, STAR:2022fan, Zhao:2024ipr, Sheng:2024kgg}, the polarization state of a virtual photon can be measured through the dilepton spectrum, especially the angular distribution~\cite{NA60:2008iqj}.

As effectively a massive spin-1 particle, the polarization states of the virtual photon are captured by a spin density matrix. With respect to the polarization vector, $\e^\mu_\lambda(Q)$, with helicity $\lambda=-1, 0, +1$, the identity $\sum_{\lambda}\e^\mu_\lambda(Q) (\e^\nu_{\lambda}(Q))^*=-g^{\mu\nu}+Q^\mu Q^\nu/Q^2$ allows for a decomposition into spin states,
\be
\label{eq:LMexp}
L^{\mu\nu}(Q_1,Q_2) W_{\mu\nu}(Q) = \sum_{\lambda \lambda'} (\e^\beta_{\lambda})^*L_{\alpha\beta}(Q_1,Q_2)\e^\alpha_{\lambda'}
(\e^\mu_{\lambda'})^* W_{\mu\nu}(Q) \e^{\nu}_{\lambda}
\equiv \sum_{\lambda,\lambda'}\rho_{\lambda,\lambda'}^{\rm lep}(Q_1,Q_2)  \rho_{\lambda',\lambda}^{\gamma}(Q)\,,
\ee
where one defines,
\be
\rho_{\lambda,\lambda'}^{\rm lep}(Q_1,Q_2) \equiv (\e^\beta_{\lambda})^*L_{\alpha\beta}(Q_1,Q_2)\e^\alpha_{\lambda'}\quad{\rm and}\quad
\rho_{\lambda,\lambda'}^{\gamma}(Q) \equiv (\e^\mu_{\lambda'})^* W_{\mu\nu}(Q) \e^{\nu}_{\lambda}\,,
\ee
as the spin density matrices of the lepton pair and the virtual photon, respectively. Note that the decomposition is acheived owing to the vector current conservation, which guarantees $Q_\mu L^{\mu\nu} = Q_\mu W^{\mu\nu}=0 $, and $Q_\mu \e^\mu_\lambda (Q)=0$.

With respect to the helicity $\lambda=\pm1$ and $0$, $\e^\mu_\lambda(Q)$ indicates the eigenstate of a massive virtual photon with transversal polarizations and longitudinal polarization, respectively. Correspondingly, as the current-current correlator being decomposed into the spin states\footnote{
Note that the current-current correlator $W^{\mu\nu}$ characterizes an intermediate vector state, namely, the virtual photon, inside a QGP medium.
}, the spin density matrix of the virtual photon, $\rho^\gamma$, contains information on polarization of the intermediate state during the di-lepton production.  Because statistical factors arise due to the thermal nature of the QGP medium, as well as the average over initial and final states during quark-antiquark annihilation, %dilepton production, 
the density matrix $\rho^\gamma$ captures a mixed state, so that $\tr (\rho^\gamma)^2<\tr  \rho^\gamma\equiv \rho_{-1,-1}^\gamma+\rho_{0,0}^\gamma + \rho_{+1,+1}^\gamma$. Here, the trace is taken in the spin subspace. 
By construction, the spin density matrix is hermitian. Therefore, upon a normalization factor which will be shown proportional to the dilepton production rate, and the hermiticity condition, the virtual photon spin density matrix $\rho^\gamma$ contains five independent components.

The differential production rate of a dilepton pair from QGP can be described in terms of the polarization states of the virtual photon. In particular, one notices that the dilepton production rate is proportional to the trace 
%$\tr \rho^{\gamma}$ 
of the virtual photon spin density matrix. This can be shown through integration with respect to the relative momentum $Q'=Q_1-Q_2$ in \Eq{eq:LMexp}, which leads to
$
%\frac{d R}{d^4 Q} \propto
\int d^4 Q' \rho^{\rm lep}_{\lambda,\lambda'} 
\propto
\int d^4 Q' \e^{\alpha}_\lambda(Q)[Q^2 g_{\alpha\beta} - Q_\alpha Q_\beta + {Q'}_\alpha {Q'}_\beta]\e^{\beta}_{\lambda'} (Q)
\propto \delta_{\lambda,\lambda
}\,,
$
and correspondingly, 
%\ly{check here: factor differs from eq. 3.51 in Speranza's thesis.}\textcolor{blue}{MH: eq.3.52? We can compare it with eq.11 in Speranza's PLB2018 paper}
\be
\label{eq:ratetrace}
\frac{d R}{d^4 Q}= 
\frac{\alpha_{\rm EM}^2}{3 \pi^3 M^2} \left(1+\frac{2m_l^2}{M^2}\right)
\sqrt{1-\frac{4 m_l^2}{M^2}}
\tr \rho^\gamma\,.
%\int d^4 Q'\sum_{\lambda,\lambda'}\rho_{\lambda,\lambda'}^{\rm lep}\rho_{\lambda',\lambda}^\gamma = \sum_\lambda \rho_{\lambda,\lambda}^\gamma\,.
\ee
\Eq{eq:ratetrace} implies that the two transversal and one longitudinal polarization states of the virtual photon contribute equally to the production of the dilepton pair, which can be understood, for instance, in an eigen system that diagnolizes the density matrix. % with $\rho_{\lambda,\lambda}^\gamma$ the eigenvalue of each eigenstate. 

%Off-diagnol 
All the individual components in the virtual photon density matrix are involved in the angular distribution of the dilepton pair. Unlike the rate, the angular distribution of the produced dileptons is not Lorentz invariant. For convenience, one chooses the virtual photon rest frame with $Q=(M,{\bf 0})$ and $Q'=(0,{\bf q}')=(0,2{\bf q}_{1})$, where $|\mathbf{q}_{1}|=|\mathbf{q}_{2}|=\sqrt{\frac{M^{2}}{4}-m_{l}^{2}}$ is the magnitude of the 3-momentum of the lepton in the virtual photon rest frame. Accordingly, given a specified quantization axis, one is allowed to parameterize the lepton four-momentum in terms of the solid angle $\Omega_{l}=(\theta_{l},\phi_{l})$,
\begin{subequations}\label{eq:leptonmomentumHX}
\begin{align}
&Q_1^\mu=\left(\sqrt{m_{l}^{2}+\mathbf{q}_{1}^{2}},|\mathbf{q}_{1}| \sin \theta_{\ell} \cos \phi_{\ell}, |\mathbf{q}_{1}| \sin \theta_{\ell} \sin \phi_{\ell}, |\mathbf{q}_{1}| \cos \theta_{\ell}\right)\,,\\
&Q_2^\mu=\left(\sqrt{m_{l}^{2}+\mathbf{q}_{1}^{2}},-|\mathbf{q}_{1}| \sin \theta_{\ell} \cos \phi_{\ell}, -|\mathbf{q}_{1}| \sin \theta_{\ell} \sin \phi_{\ell}, -|\mathbf{q}_{1}| \cos \theta_{\ell}\right)\,.
\end{align}
\end{subequations}
Substituting \Eqs{eq:leptonmomentumHX} into \Eq{eq:lmunu} yields an expression of $L^{\mu\nu}$ in terms of $\theta_{\ell}$ and $\phi_{\ell}$~\cite{Speranza:2016tcg}, and accordingly leads to~\cite{Gottfried:1964nx, Schilling:1969um,Faccioli:2011pn},
%for a re-expression of the angular dependent production rate,
%\ly{these vectors can be specified later}
%In the virtual photon rest frame, the polarization vectors $\e_\lambda^\mu(Q)$ can be chosen as
%\begin{equation}
%\label{eq:pol-vector}
%\begin{aligned}
%\e^\mu_{\pm 1}  =\mp \frac{1}{\sqrt{2}}(0,1,\pm i, 0), \qquad
%\e^\mu_0  =(0,0,0,1), 
%%\\
%%\e^\mu_{+1} & =-\frac{1}{\sqrt{2}}(0,1, i, 0) .
%\end{aligned}
%\end{equation}
%such that the virtual photon is circularly polarized in the transverse direction. 
%With these parameterizations, the production rate of dilepton pair with angular dependence is found to be
%\begin{equation}
\begin{align}
\label{eq:dRangular}
\frac{d R}{d^4 Q d \Omega_{\ell}} = \frac{\alpha_{\rm EM}^2}{16 \pi^4 M^2}
\left(1-\frac{4 m_l^2}{M^2}\right)^{3/2}
%\sum_{\lambda,\lambda'}\rho_{\lambda,\lambda'}^{\rm lep}\rho_{\lambda',\lambda}^\gamma  =
\mathcal{N}(1&+ \lambda_\theta \cos ^2 \theta_{\ell} 
 +\lambda_\phi \sin ^2 \theta_{\ell} \cos 2 \phi_{\ell} + \lambda_\phi^{\perp} \sin ^2 \theta_{\ell} \sin 2 \phi_{\ell}\nonumber\\
&+\lambda_{\theta \phi} \sin 2 \theta_{\ell} \cos \phi_{\ell}+\lambda_{\theta \phi}^{\perp} \sin 2 \theta_{\ell} \sin \phi_{\ell})\,,
\end{align}
%\end{equation}
where $\mathcal{N}=(1+2m_l^2/|{\bf q}_{1}|^2)\tr \rho^\gamma + \rho_{0,0}^\gamma$ is the normalization factor. These five independent anisotropic coefficients $\lambda_\theta$, $\lambda_\phi$,  $\lambda_{\theta\phi}$, $\lambda_{\phi}^{\perp}$ and $\lambda_{\theta\phi}^\perp$, are defined through the components of the spin density matrix of the virtual photon. For instance,
\be
\lambda_{\theta} = \frac{1}{\mathcal{N}}(\rho^\gamma_{-,-} + \rho^\gamma_{+,+} - 2 \rho^\gamma_{0,0})\,,\quad
\lambda_{\phi}  = \frac{2}{\mathcal{N}} {\rm Re} \rho^\gamma_{+,-} \,,
\quad
\lambda_{\theta\phi} = \frac{\sqrt{2}}{\mathcal{N}} {\rm Re}(\rho^\gamma_{+,0}-\rho^\gamma_{-,0})\,.
\ee

\begin{figure}
    \centering
    \includegraphics[width=0.5\textwidth]{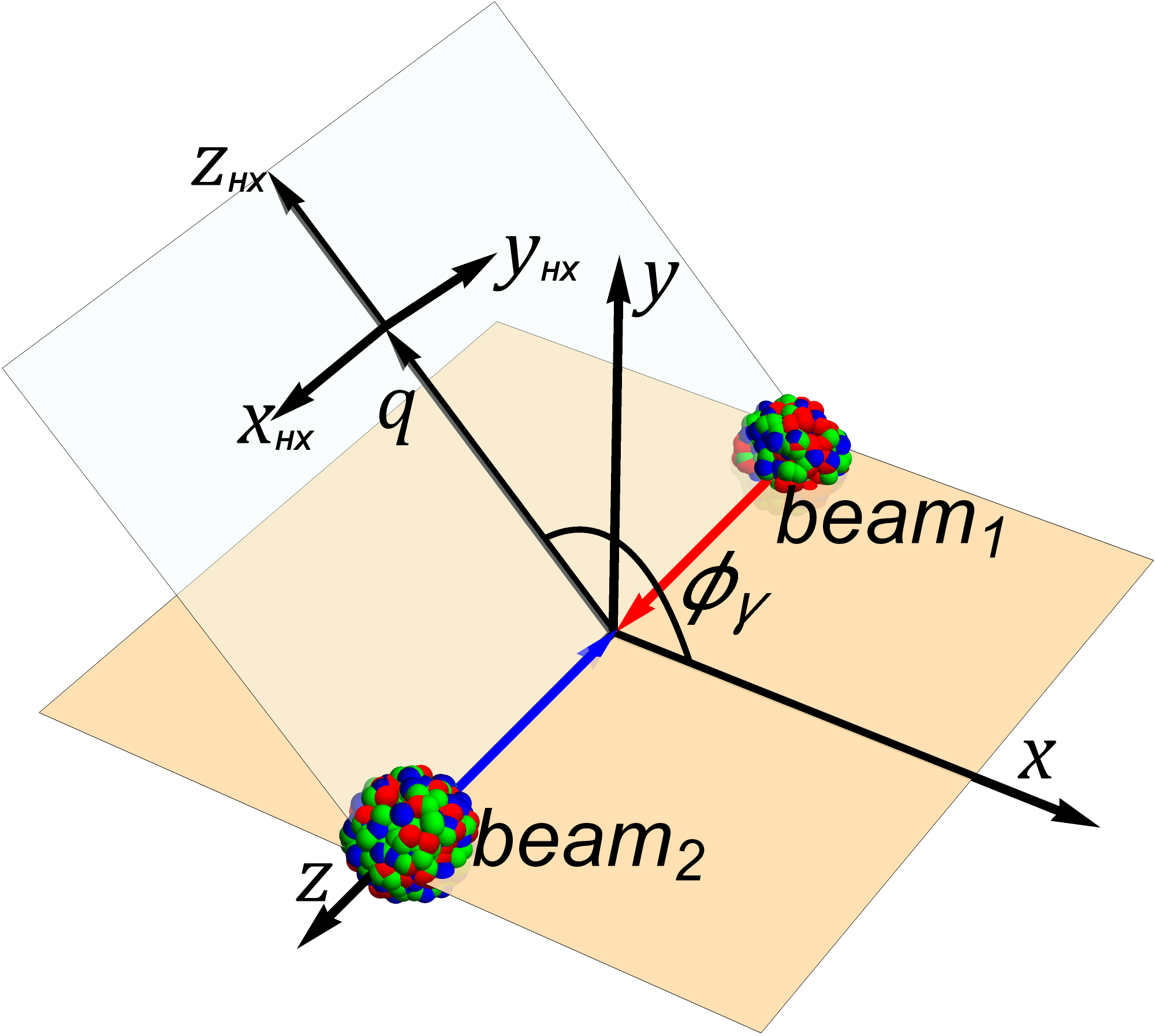}
    \caption{Coordinate configuration in the helicity frame. As the quantization axis ($Z_{HX}$-axis) is given by the three momentum of the virtual photon, $\q$, the production plane is then determined by $Z_{HX}$ and the beam axis ($z$-axis), which accordingly determines $X_{HX}$. And $\phi_{\gamma}$ is the angle between the virtual photon and the reaction plane}
    \label{fig:HXframe}
\end{figure}

Several comments are in order. First, in the dilepton angular distribution, the six independent angular modes (including the constant term) in \Eq{eq:dRangular} are originated from the $Q'_\alpha Q'_\beta$ factor in the lepton tensor. 
In the local rest frame of the virtual photon, for instance, the factor reduces to the direct product of three-vectors, $q'_i q'_j$, which can be decomposed into six independent irreducible tensors with respect to the SO(3) symmetry group. In fact, the independent number of angular modes should be identical %six 
in arbitrary frames, owing to the transversality condition $Q\cdot Q'=0$.\footnote{
For the electromagnetic decay of a virtual photon, this transversality condition is a consequence of the conservation of lepton number. However, for the weak decay process of a vector meson, the transversality condition is not available and hence one expects more angular modes in the angular distribution of final state particles.
}
Secondly, even in the virtual photon rest frame, these anisotropic coefficients depend on the specification of a quantization axis. Throughout this work, we choose the so-call helicity frame (HX), in which the quantization axis follows the direction of the three momentum of the virtual photon,
$\hat z_{\rm HX}=\hat q$.
A production plane, which consists of the beam axis in experiments ($z$-axis) and the quantization axis, can be determined accordingly.  The helicity frame is feasible in realistic experiments, as long as the three momentum of generated dilepton pairs is identified. The configuration of the helicity frame coordinates is depicted in \Fig{fig:HXframe}.  
For later convenience, we also choose in the helicity frame the polarization vectors (circular polarization) as,
\begin{equation}
\label{eq:pol-vector}
\begin{aligned}
\epsilon^\mu(-1) & =\frac{1}{\sqrt{2}}(0,1,-i, 0), \\
\epsilon^\mu(0) & =(0,0,0,1), \\
\epsilon^\mu(+1) & =-\frac{1}{\sqrt{2}}(0,1, i, 0) .
\end{aligned}
\end{equation} 
Thirdly, these coefficients reveal the nature of virtual photon polarization. For instance, in the limit $m_l\to 0$, one notices that $\lambda_\theta$ reduces to,
\[
\lambda_\theta = \frac{\rho^\gamma_{-,-} + \rho^\gamma_{+,+} - 2 \rho^\gamma_{0,0}}{\rho^\gamma_{-,-} + \rho^\gamma_{+,+} + 2 \rho^\gamma_{0,0}} \in [-1,1]\,,
\]
where the lower bound ($-1$) is realized with respect to a longitudinally polarized virtual photon with $\rho^\gamma_{\pm,\pm} = 0$, and the upper bound ($+1$) corresponds to a transversely polarized virtual photon with $\rho^\gamma_{0,0}=0$. For an unpolarized virtual photon, namely, $\rho^\gamma_{-,-}=\rho^\gamma_{+,+} = \rho^\gamma_{0,0}$, $\lambda_\theta$ vanishes.
The angular modes involving $\phi_{\ell}$ break azimuthal symmetry in the virtual photon rest frame, reflecting the mixing among polarization states of the virtual photon.

\subsection{Polarization states of a virtual photon inside a weak magnetic field}
\label{sec:polarB}

Being generated from the in-medium Drell-Yan process, the polarization states of a virtual photon depend strongly on the motion of quarks. % inside the QGP. 
In a static medium, for instance, where the quarks are distributed in phase space isotropically, in the rest frame of the virtual photon a longitudinal polarization could be induced owing to the Lorentz boosted flow velocity $U^\mu=\gamma(1,-\vec v_{\gamma*})$~\cite{NA60:2008iqj, Speranza:2018osi}. Moreover, an even polarized state should be expected in the medium in the presence of an extra vector that breaks the rotational symmetry. For instance, 
%a non-trivial flow velocity, $U^\mu$, during the expansion of QGP fireball, 
a finite electromagnetic field  in the fluid local rest frame 
gives rise to the electric field, $E^\mu = F^{\mu\nu} U_\nu$, where $F^{\mu\nu}$ is the electromagnetic field strength tensor.
%\footnote{
%Rotational symmetry can also be broken due to the medium expansion with a non-trivial flow velocity.
%}. 

In the case of a weak magnetic field, the hydrodynamic description of QGP contains dissipative correction associated with the conserved electric charge. In terms of the net electric charge flow, it is $\Delta j^\mu = \sigma_{\rm el} E^\mu$, where $\sigma_{\rm el}$ is the electrical conductivity of the QGP. In a kinetic theory approach, correspondingly, the dissipative correction to quark local equilibrium distribution can be solved. To the leading order of $|eB|/T^2$, one finds for quark species $a=u,d,s$ and $\bar u, \bar d, \bar s$, with electric charge number $Q_a$~\cite{Sun:2023rhh},
\be
\label{eq:dfem}
\delta f^{(a)}_{\rm EM}(X,\p) = \frac{n_\eq(1+n_\eq)}{P\cdot U} 
\frac{\sigma_{\rm el}}{T \chi_{\rm el}} eQ_a F^{\mu\nu} P_\mu U_\nu\,,
\ee
where $\chi_{\rm el}$ is an effective charge susceptibility of the QGP system. 
Note that in \Eq{eq:dfem} the spin degrees of freedom of individual quarks have been averaged.
More detailed information regarding \Eq{eq:dfem} can be found in \cite{Sun:2023rhh}.
In the current study, we shall ignore the dissipative correction due to viscosities, so that the in-medium average of the virtual photon polarization tensor is given by an integral in \Eq{eq:emsemble}, with
$f_q = n_\eq + \delta f_{\rm EM}$.
%\footnote{
%Because viscous corrections in the QGP medium dose not involve a vector that breaks the isotropy of quark phase space distribution in the fluid local rest frame, 
%}. 
\begin{figure}
    \centering
    \includegraphics[width=0.35\textwidth]{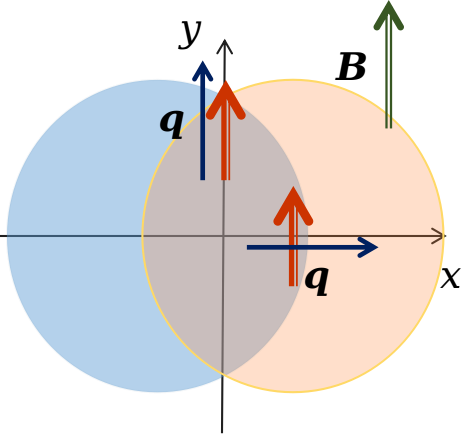}
    \caption{(Color online) Thermal virtual photon polarization (red arrows) is induced aligned (or anti-aligned, not shown) with the external magnetic field (green arrow) in heavy-ion collisions. Correspondingly, the emitted virtual photons along the $x$-axis are expected to be transversely polarized, while the virtual photons emitted out of the reaction plane are expected to be longitudinally polarized. }
    \label{fig:polarB}
\end{figure}

In the lab frame, as the QGP being created by the colliding nuclei, a finite electromagnetic field is also generated by the relativistic motion of spectators, especially in non-central collisions\footnote{
Even in ultra-central collisions, on an event-by-event basis, the external electromagnetic field presents with a random orientation, although the field is expected to vanish over the event average. 
}. 
Up to fluctuations, the induced electromagnetic field is dominated by its magnetic component, orientated out of the reaction plane, namely, ${\bf B} = B_y \hat y$. Accordingly, \Eq{eq:dfem} implies that the dissipative contribution to virtual photons via the weak external magnetic field %induces virtual photons from
involves quarks with a finite velocity in the reaction plane ($x$-$z$ plane). Since the quark spin has been averaged, as the quark-antiquark pair annihilates, the polarization states of the produced virtual photons are related to the quarks' orbital angular momentum, which is aligned or anti-aligned with the magnetic field. Therefore, as demonstrated in \Fig{fig:polarB}, in the presence of a weak external magnetic field, in the lab frame the virtual photons are polarized along the magnetic field. Correspondingly, the virtual photons emitted along the $x$-axis are transversely polarized, while the virtual photons emitted out of reaction are longitudinally polarized.

\section{Dilepton polarization from the Bjorken flow}
\label{sec:numericalresults}

As shown in \Eq{eq:dfem}, the coupling of the weak magnetic field to QGP requires a finite flow velocity. To this end, %In order to verify the effect of weak magnetic field on the virtual photon polarization, 
we consider, for the simplest case, the Bjorken flow, which has a longitudinal expansion. 

Bjorken flow gives rise to an analytical description of the 1+1 dimensional expanding QGP,\footnote{
Bjorken flow is actually 0+1 dimensional since the dependence on space-time rapidity in the Milne coordinates is suppressed.
}
subject to the Bjorken boost symmetry along the longitudinal direction determined by the collision axis, and the translational invariance in the transverse plane. It is believed that Bjorken flow approximates the early-stage evolution of QGP in high-energy heavy-ion collisions, during which the medium expansion is dominated in the longitudinal direction until a time scale that is comparable to the system size. 
On the other hand, one also notices that owing to the temperature dependence of the dilepton production rate, there are more thermal dileptons generated during the early times of the QGP evolution, which makes our calculation more reliable with respect to the realistic high-energy heavy-ion collisions.

In the Milne coordinates, where the proper time and the space-time rapidity are introduced respectively as,
\be
\tau = \sqrt{t^2-z^2}\,,\qquad
\xi = \tanh^{-1} \left(\frac{z}{t}\right)\,,
\ee
the flow four velocity in Bjorken flow is well determined, $u^\mu=(1,0,0,0)$, and
the medium evolution depends only on $\tau$. For instance, as we shall consider for simplicity in our calculation, the evolution of temperature in an ideal fluid is given as
\be
T(\tau) = T(\tau_0) \left(\frac{\tau_0}{\tau}\right)^{1/3}\,.
\ee
This information suffices to allow one to evaluate \Eq{eq:dfem} and to calculate the production of the thermal dileptons. We thus follow all the procedures introduced in the previous sections, and focus on the electron-position pair production, with electron mass  $m_{e}=0.51$ MeV. For quarks in the QGP medium, our analysis includes up and down quarks and their anti-particles, with their masses $m_{u}=m_{d}=m_{q}=5$ MeV.

In high-energy heavy-ion collisions, the external magnetic field decays as the QGP medium expands. Due to the lack of knowledge of the electrical conductivity in QGP, however, the exact space-time dependence of the magnetic field remains undetermined so far. In this work, we consider a magnetic field that distributes homogeneously in space in the lab frame, while its time dependence is given by~\cite{Shi:2017cpu, Xu:2020sui, Huang:2022qdn},
\be
\label{eq:Bdecay}
B(\tau) = \frac{B_0}{1+(\tau/\tau_B)^2}\,.
\ee
In the above equation, $B_0$ corresponds to the field strength at the instant of the nucleus-nucleus collision, and the parameter $\tau_B$ is used to characterize the lifetime of the magnetic field. With respect to the mid-central Au-Au collisions with $\sqrt{s_{\rm NN}}$ = 200 MeV at RHIC, we take $e B_0=3 m_\pi^2$~\cite{Skokov:2009qp,  Deng:2012pc}. The lifetime of the magnetic field is to a large extent determined by the appearance of a Lorentz factor in the relativistic collisions, and a small enhancement due to the finite electrical conductivity in the medium~\cite{Inghirami:2016iru, Huang:2022qdn, Yan:2021zjc}. Nonetheless, at the top RHIC energies, the lifetime of the magnetic field is expected no longer than $1$ fm/c. In this work, we shall allow $\tau_B$ to increase from $0.2$ fm/c, to $0.4$ fm/c and $0.6$ fm/c, correspondingly, the effect of the field on the QGP evolution gets stronger. 

The medium properties of QGP are determined by the solution of the ideal Bjorken flow. We take $T(\tau_0)= 300$ MeV at $\tau_0=1.0$ fm/c, and sum up the produced dileptons down to $\tau_f = 5$ fm/c. We choose a temperate dependent electrical conductivity, with its value extracted from perturbative quantum chromodynamics calculation, $\sigma_{\rm el}/T=2$~\cite{Arnold:2003zc}.

For convenience, we do not distinguish the difference between the center-of-momentum frame and the lab frame. In the lab frame,  the momentum of the virtual photon can be expressed by:
\begin{equation}
\left(Q^\mu\right)_{\mathrm{Lab}}=\left(M_T \cosh y, q_T \cos \phi_\gamma, q_T \sin \phi_\gamma, M_T \sinh y\right)
\end{equation}
where $M_{T}=\sqrt{q_{T}^{2}+M^{2}}$ is the transverse mass and $q_{T}$ is the transverse momentum of dilepton. $\phi_{\gamma}$ is the angle between the dilepton(virtual photon) and the reaction plane, and $y$ is the rapidity. We also use the $Q_{\perp}=\left(q_T \cos \phi_\gamma, q_T \sin \phi_\gamma\right)$ to make the following expressions compact.

\begin{figure}
    \centering
            \includegraphics[width=0.45\textwidth]{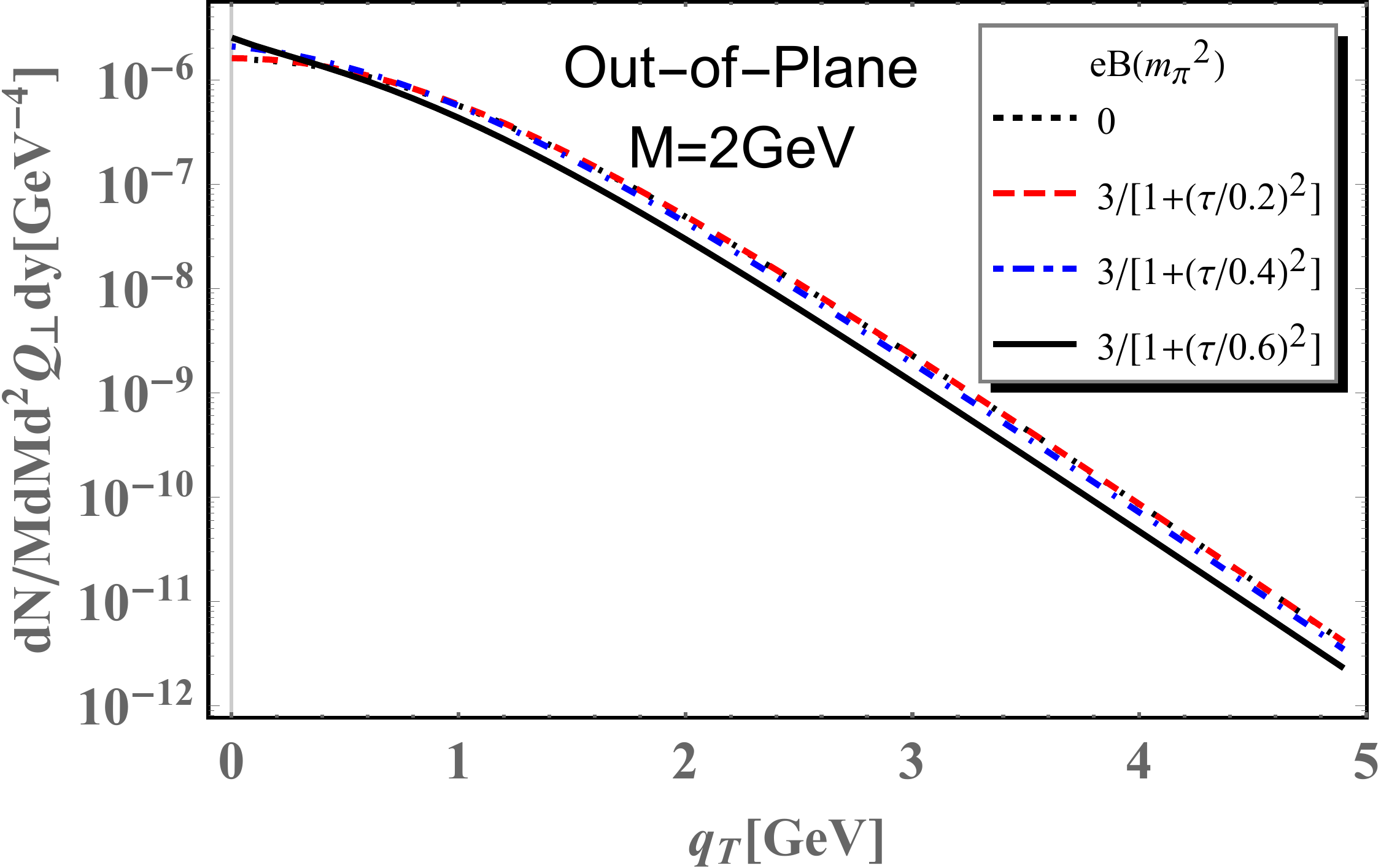}
                \includegraphics[width=0.45\textwidth]{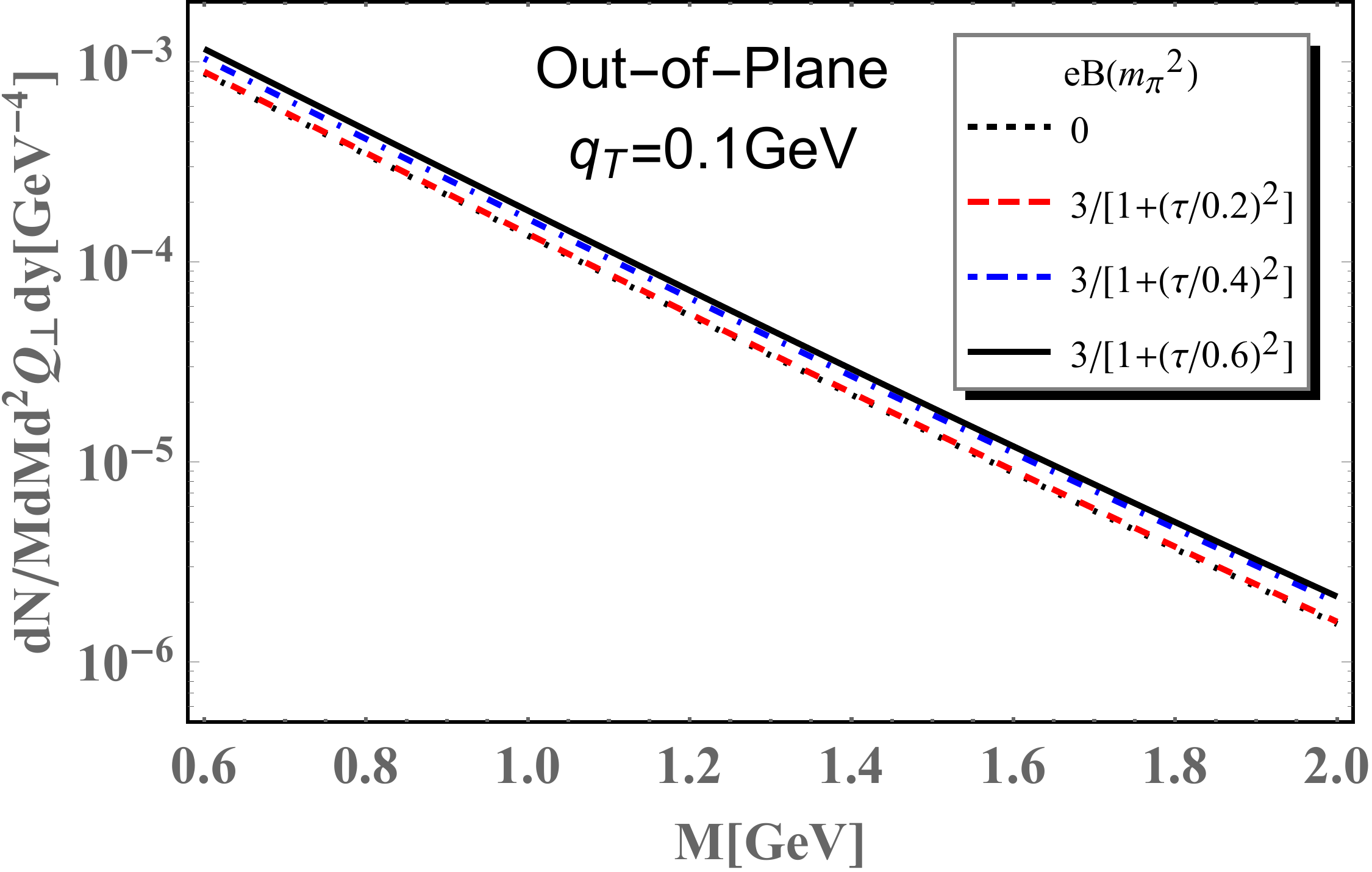}
    \caption{(Color online) The production rate of the electron-positron pair as a function of the virtual photon transverse momentum $q_T$ (left) and invariant mass $M$ (right). The rapidity is fixed at $y=0$. The black dotted lines stand for zero magnetic field case. The red dashed, blue dot-dashed, and black solid lines stand for finite magnetic fields with lifetime $\tau_{B}=0.2,0.4,0.6$ fm/c, respectively.
    }
    \label{fig:rate}
\end{figure}

\begin{figure}
    \centering
            \includegraphics[width=0.46\textwidth]{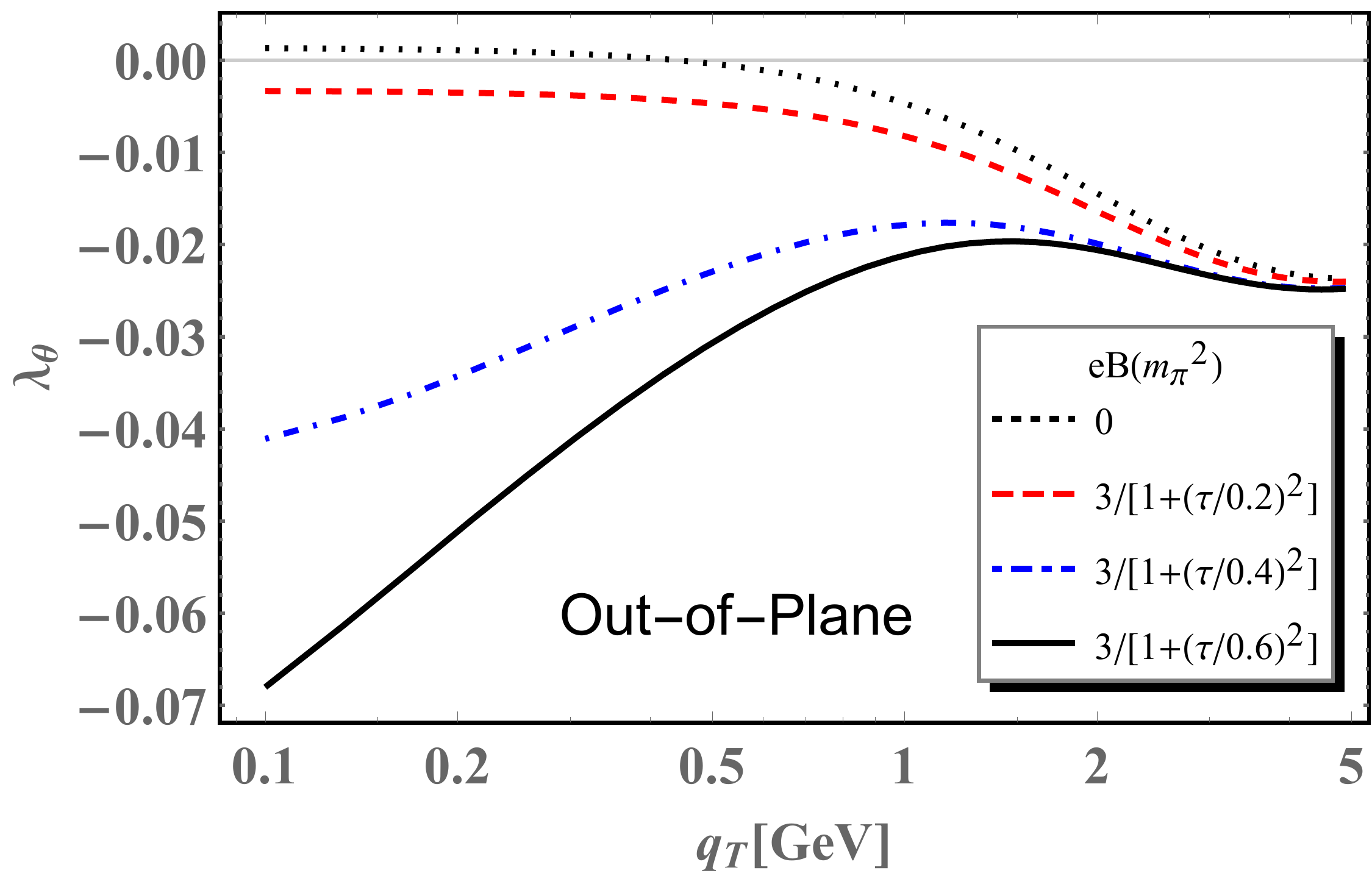}
                \includegraphics[width=0.45\textwidth]{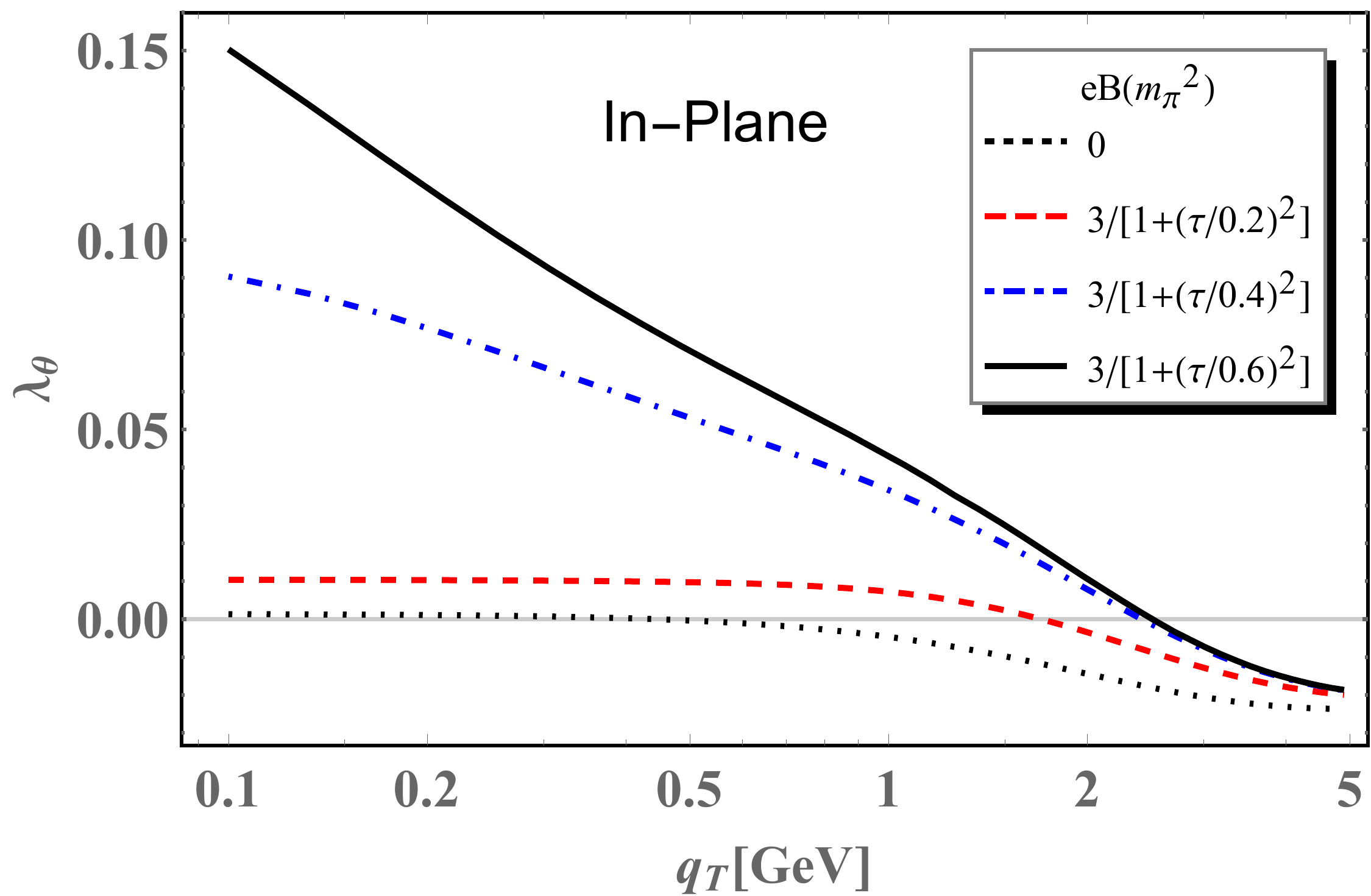}
    \caption{$\lambda_{\theta}$ as a function of transverse momentum $q_{T}$ from QGP with Bjorken expansion. The black dotted lines stand for zero magnetic field case. The red dashed, blue dot-dashed, and black solid lines stand for finite magnetic fields with lifetime $\tau_{B}=0.2,0.4,0.6$ fm/c, respectively.}
    \label{fig:ltheta}
\end{figure}

In \Fig{fig:rate}, the differential production rate of the thermal dilepton in QGP is shown as a function of the transverse momentum of the virtual photon $q_T$ with fixed virtual photon invariant mass $M=2$ GeV (left panel), and a function of the virtual photon invariant mass $M$ with $q_T=0.1$ GeV (right panel). In comparison to the case of pure QGP evolution (black dotted lines), the weak external magnetic field leads to small corrections to the dilepton spectrum.

With respect to the dilepton polarization, on the other hand, the effect of the weak magnetic field is remarkable, as shown through the anisotropic coefficient $\lambda_\theta$ in \Fig{fig:ltheta}. Here, we take the invariant mass as $M=2$ GeV, at which the dominant non-cocktail contribution to the dilepton spectrum is from QGP~\cite{PHENIX:2015vek}. 
In previous studies~\cite{Speranza:2018osi}, it has been noticed that, due to the Lorentz boost, even for a QGP with Bjorken expansion, the produced dileptons are polarized with a finite $\lambda_\theta$. Such an effect is reproduced and shown as the black dotted lines in \Fig{fig:ltheta}. Because the background QGP medium is azimuthally symmetric, without the external magnetic field, the resulted dilepton polarization is azimuthally symmetric as well with respect to the emission angle of the dilepton, $\phi_\gamma$ (cf. \Fig{fig:HXframe}). In the presence of a weak external magnetic field, the polarization in the produced dilepton gets enhanced significantly. Moreover, the polarization of the dilepton reflects apparent azimuthal angle dependence. In the small $q_T$ region, where the effect of the magnetic field overwhelms the effect of Lorentz boost, one observes longitudinal polarization for the out-of-plane dileptons ($\phi_\gamma=\pi/2$), and transverse polarization for the in-plane dileptons ($\phi_\gamma=0$), in consistency with our previous analysis in \Sect{sec:polarB}. In the large $q_T$ region, the polarization in the dilepton is dominated by the Lorentz boost, hence the dependence on the magnetic field becomes negligible. 

\begin{figure}
    \centering
    \includegraphics[width=0.45\textwidth]{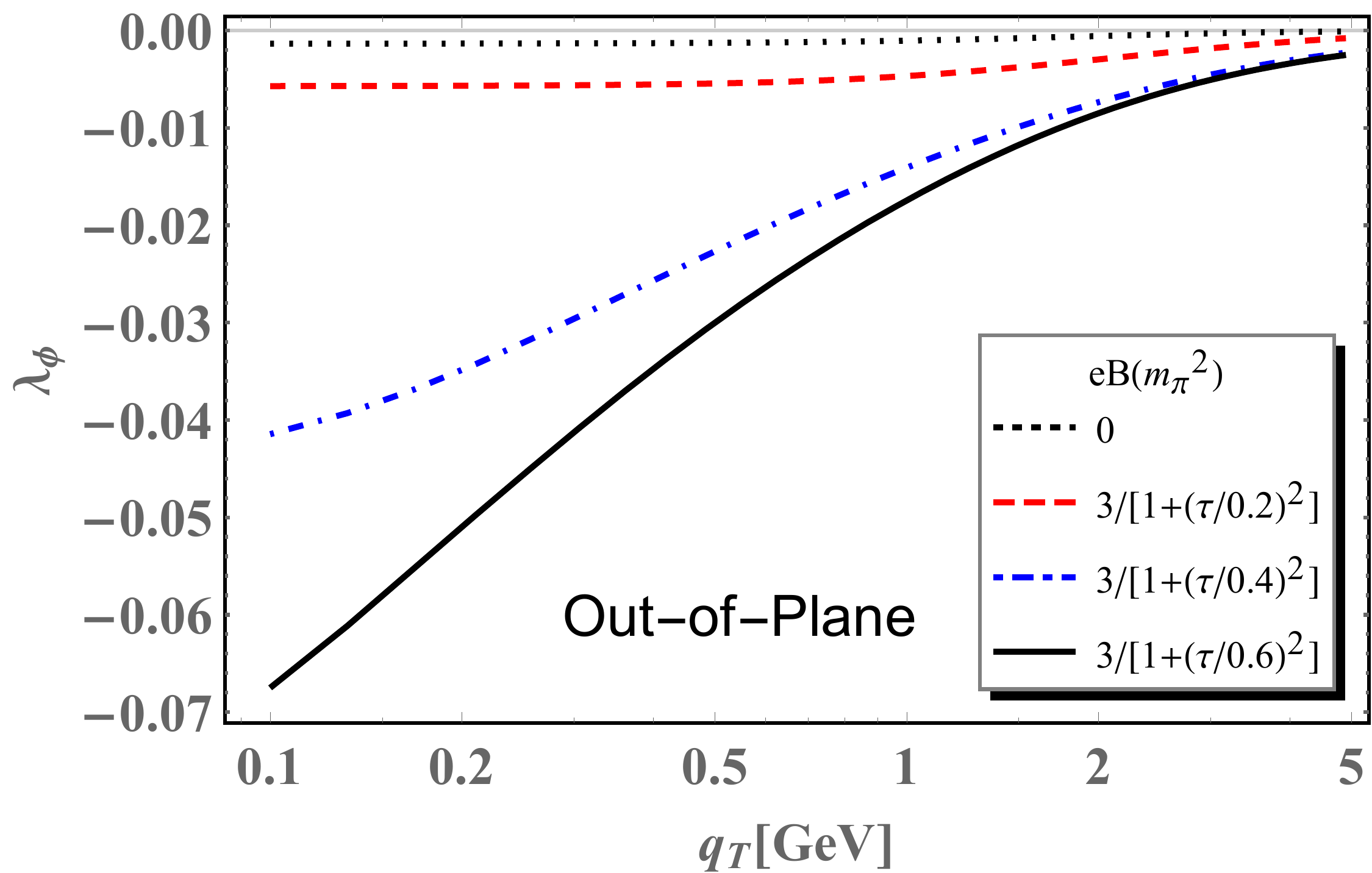}
        \includegraphics[width=0.47\textwidth]{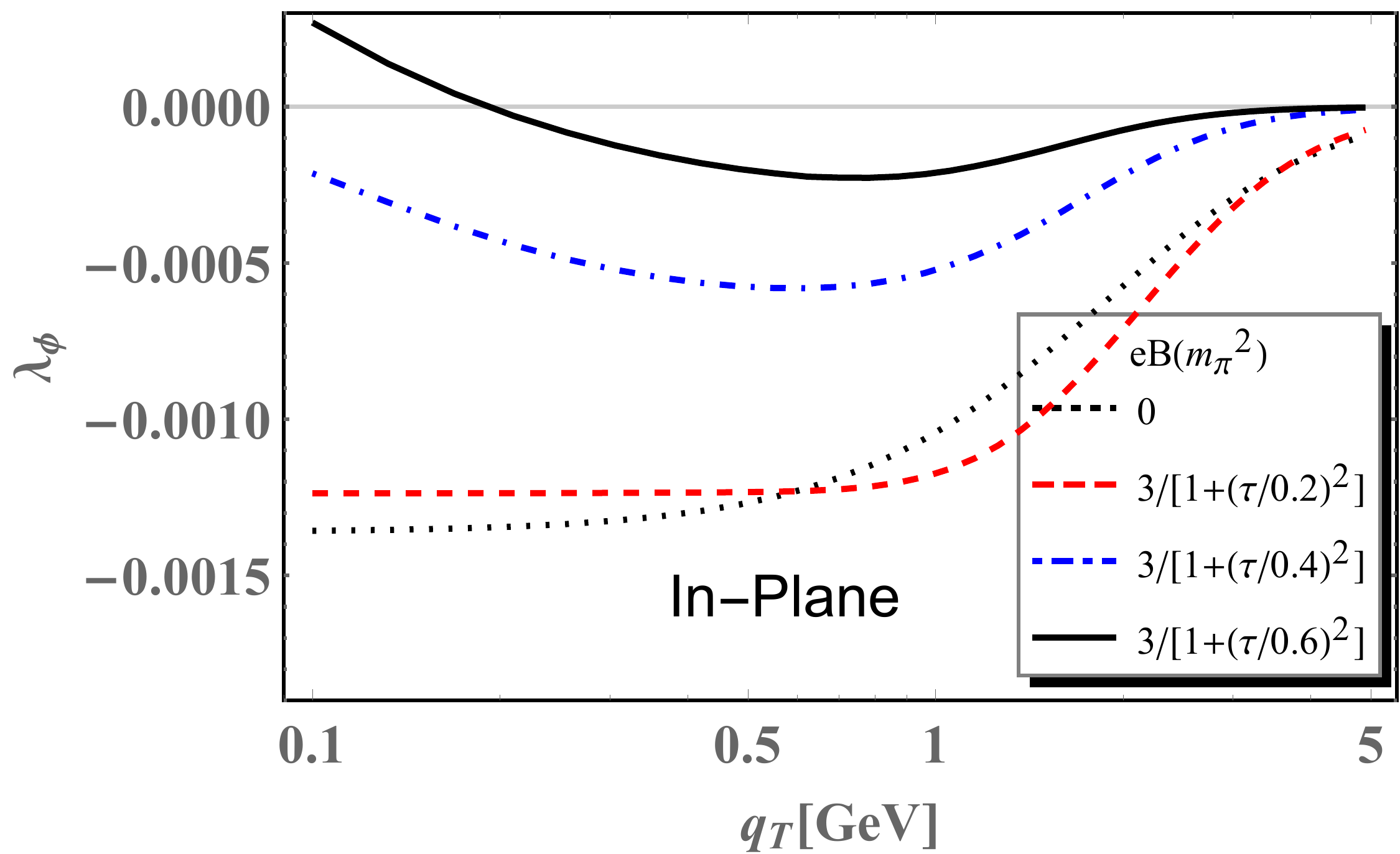}
    \caption{$\lambda_{\phi}$ as a function of transverse momentum $q_{T}$ in a Bjorken flow.The black dotted lines stand for zero magnetic field case. The red dashed, blue dot-dashed, and black solid lines stand for finite magnetic fields with lifetime $\tau_{B}=0.2,0.4,0.6$ fm/c, respectively.}
    \label{fig:lphi}
\end{figure}

Regarding the dileptons with invariant mass $M=2$ GeV, we also calculated $\lambda_\phi$, with results shown in \Fig{fig:lphi}. Because the coefficient $\lambda_\phi$ characterizes the polarization in the azimuthal direction with respect to the virtual photon momentum, the effect from the Lorentz boost is marginal. However, in the presence of the weak external magnetic field along the $y$-axis, Bjorken flow implies the coupling of the magnetic field to quarks moving in the reaction plane, namely, $B_y u_z p_x$. As a consequence, for the di-lepton pairs produced along the $x$-axis, or in plane, azimuthal symmetry with respect to the virtual photon three momentum is barely broken, and the induced $\lambda_\phi$ is small. On the other hand, for the di-leptons emitted out of the reaction plane, the azimuthal symmetry with respect to the three momentum is apparently broken, one accordingly gets a finite and negative $\lambda_\phi$, as shown in the figures.

\section{Summary and discussion}
\label{sec:sum}

In this work, we analyzed the polarization property of thermal dileptons from QGP. Especially, we found that an external magnetic field in the QGP medium, albeit weak, could induce a significant effect on the polarization of the produced dileptons. Quite similar to the direct photon production and lambda hyperon polarization~\cite{Sun:2023rhh, Sun:2024isb}, the weak magnetic effect in QGP, has a small correction to the transverse momentum and invariant mass dependence of the dilepton spectrum. However, as the motion of quark and anti-quark pairs get extra alignment associated with the orientation of the magnetic field, the resulting change in the dilepton angular distribution can be remarkable. 

Using the ideal Bjorken flow to describe the QGP expansion, we calculated the anisotropic coefficients in the dilepton spectrum. Although the calculation is simplified, for instance, medium expansion in the transverse directions and anisotropic flow of the QGP are neglected, to a large extent, we believe our results are reliable. First, as we mentioned previously, the emission of dileptons from QGP is dominated in the early stages, where the medium expansion is mainly one dimensional. Secondly, at the early stages, the anisotropic flow of the QGP is not substantial, its influence to the dilepton production is expected small. 

Our results indicate that the anisotropic coefficients in the dilepton spectrum, $\lambda_\theta$ and $\lambda_\phi$ are both observables sensitive to the external magnetic field. For the low $p_T$ emission and invariant mass $M=2$ GeV, these coefficients reach up to $10^{-1}$, which we expect to be measurable at the RHIC collisions within errors.  In particular, the dependence on the emission angle provides a qualitative probe to identify the existence of the magnetic field in relativistic heavy-ion collisions. As a major result of the current study, we propose that a finite and positive $\lambda_\theta$, and a negligible $\lambda_\phi$, for the dileptons emitted in the reaction plane, together with the finite and negative $\lambda_\theta$ and $\lambda_\phi$ for the dileptons emitted out of the reaction plane, as the unambiguous signature of the effect of the external magnetic field.

\begin{acknowledgments}
We would like to thank Qiye Shou for the very helpful discussion that motivated our study. 
We wish to acknowledge the support of the Natural Science Foundation of Shanghai (No. 23JC1400200). This work is also supported partly by the National Natural Science Foundation of China (NSFC No. 12375133 and No. 12147101). 
\end{acknowledgments}

\bibliography{virtualGamma}% Produces the bibliography via BibTeX.

\end{document}